# Non-reciprocity of spin wave propagation induced by the interface Dzyaloshinskii-Moriya interaction in Py/Pt film structures.

A. A. Stashkevich[1], M. Belmeguenai[1],Y. Roussigné[1], S.M.Cherif[1], M. Kostylev[2], M. Gabor[3], D. Lacour[4], C. Tiusan[3,4] , M. Hehn[4]

*[1]LSPM (CNRS-UPR 3407), Université Paris 13, 93430 Villetaneuse, France*

*[2]School of Physics, M013, University of Western Australia, Crawley 6009, Western Australia, Australia*

*[3]Technical University of Cluj-Napoca, Str.Memorandumului No. 28,RO-400114 Cluj-Napoca, Romania*

*[4]Institut Jean Lamour, CNRS, Université de Lorraine, 54506 Vandoeuvre, France*

Results of a comprehensive study by means of Brillouin spectroscopy, complemented by Ferromagnetic Resonance characterization, of spin waves (SW) propagating in Py/Pt bi-layers, characterized by pronounced interface Dzyaloshinskii-Moriya interactions (IDMI) are reported. Non-conventional wave behavior of SW travelling in opposite directions, characterized by non-reciprocity with respect to the inversion of the sign of the SW wave-number, has been revealed. The value of the effective IDMI constant *D* has been estimated.

The exchange interaction directly linking adjacent magnetic atoms on the microscopic quantum level is responsible for the most spectacular manifestation of the ferromagnetic ordering: the domain structure. For several decades it was universally accepted that the exchange interactions are satisfactorily described by the isotropic Heisenberg's model proposed in the late 1920ies. However, thirty years later this fundamental item had been revisited and it was shown that in low-symmetry systems an anti-symmetric term accounting for the so-called Dzyaloshinskii-Moriya interaction (DMI) has to be added [1,2]. This symmetry reduction can occur on the microscopic level in inversion asymmetric crystal fields and considerably influence behavior of naturally formed magnetic structures, such as domain walls. Thus, DMIs is of major importance in the conversion of the inner structure of the domain wall from a Bloch type to a Néel type with a preferred chirality [3] and can induce chiral structures such as spin spirals and skyrmions [4-11]. In other words, DMI can be regarded as an additional degree of freedom in ferromagnetic behavior allowing for creating tailor-made chiral magnetic nano-structures otherwise unobtainable.

Nowadays practically all the applications of thin magnetic films, however different, are unthinkable without nano-patterning unavoidably leading to breaking of symmetry [12]. This is another means of diversifying potential configurations leading to physical effects where the role of DMIs is instrumental. One of such promising possibilities is realized via the interfacial Dzyaloshinskii-Moriya interaction (IDMI). It has been a subject of significant interest recently [13-18]. In Ref.[19] a theory, based on a microscopic approach, of spin waves in ferromagnetic atomic monolayers with IDMI was constructed. It has been found that this interaction may lead to significant non-reciprocity of the spin waves (SW) in these materials. Importantly, macroscopic formalism employed in a later paper [20] has confirmed earlier theoretical predictions of non-negligible SW reciprocity for films of finite *macroscopic* thickness (on the order from several to several tens of nanometers). These films are of great technological importance, furthermore, the larger thickness makes the experimental observation of the IDMI-induced spin wave nonreciprocity more easily implementable. In terms of general wave physics this can be interpreted as linear spatial dispersion. In optics it is responsible for optical activity in chiral crystals [21]. Not surprisingly, the *symmetric* conventional exchange interaction produces quadratic dispersion scaled as $(k_{sw})^2$, an *even* function of the SW wave-number while the anisotropy engendered by the *anti-symmetric* DM exchange is characterized by an *odd* linear functional dependence of the wave frequency on $k_{sw}$.

In the present work we have investigated the IDMI induced non-reciprocity of SW propagating in thin Permalloy (Py=$Ni_{80}Fe_{20}$) films with thickness varying from 2 nm to 10 nm. The Py ($d_{Py}$=2, 4, 6 and 10 nm)/Pt (6nm) bilayers were grown by UHV sputtering in Ar pressure of 5 mtorr' on Si substrates.

The influence of IDMI on the wave behavior of magnetic modes is implemented through additional pinning due to DM induced surface anisotropy. The role of the 6 nm Pt top layer was to boost the IDMI on the Py/Pt interface; Platinum is universally known as a material with a strong spin-orbit interaction. Another consideration of major importance not to be overlooked here is taking into account the conventional normal uniaxial surface anisotropy (NUSA) unavoidably appearing on thin magnetic films which is capable of producing non-reciprocity of its own. While interpreting our experimental results, special attention has been attached to individual estimation of these two contributions, the IDMI and conventional one. It should be noted that manifestations of these two effects are entirely different.

For the easy-axis NUSA, the respective surface anisotropy constant $k_{\mathrm{surf}}$ is positive and independent from the spin wave wave number $k_{sw}$. This means that, even if it is strong enough, its presence is observable only if the characteristic Damon-Eshbach (DE) spatial asymmetry (of the dynamic magnetization distribution across the film) is sufficiently pronounced, in other words in relatively thick films. On the contrary, the IDMI pinning is linear in $k_{sw}$ and, more importantly, is sensitive to the $k_{sw}$ sign. This means, in particular, that the sign of anisotropic magnetic field standing behind pinning is inversed with inversion of the wave number

$$\vec{H}_{conv}^{(surf)} = const \quad \text{while} \quad \vec{H}_{DM}^{(surf)}(k_{SW}) = -\,\vec{H}_{DM}^{(surf)}(-k_{SW}). \qquad (1)$$

As a result no asymmetry of DE spatial profile is needed for the sought non-reciprocity being created in relatively thin films, provided $k_{sw}$ is great enough. This issue is of fundamental significance for the rest of this paper and will be revisited later in this paper. It is exactly for this reason that we have chosen Brillouin Light Scattering (BLS) giving access to SW wave-numbers as large as 20 $\mu m^{-1}$, as principal experimental tool in this study.

In any case, the non-reciprocity of SW propagation is due to the asymmetry (with respect to the sign of $k_{sw}$) of the transverse distribution of the dynamic magnetization in a DE mode. The latter is shaped by two major factors, namely, the overall exponential decay brought about by the classical long-range dipole-dipole interactions (DDIs) and inhomogeneity due to the surface anisotropy localized near this surface. Very approximately the final asymmetry can be seen as a “product” of these two contributions.

Importantly, the IDMI contribution to the frequency of the ferromagnetic resonance (FMR, corresponds to a spin wave with $k_{sw}$=0) vanishes [20], therefore to quantify the strength of NUSA we have made a series of measurements of FMR spectra for our Py films. As is well known [22] the effective field of the conventional normal uniaxial surface anisotropy NUSA scales as the inverse of the film thickness $d_{Py}$. If the anisotropy is of the easy-axis ($k_{surf} > 0$) type, this field reduces the effective value of the saturation magnetization. In Fig.1 are presented the effective magnetization ($4\pi M_{eff}$) values extracted from the best fit of the in-plane field dependences of the FMR resonance frequencies as a function of 1/ $d_{Py}$ The constant of NUSA thus obtained is equal to $k_{surf}$=0.45 erg/cm$^2$. Unfortunately, with a film as thin as this it is impossible to specify the contribution of each particular surface to this overall figure, but the sign of the slope of this dependence suggests that this anisotropy is of the easy-axis type which is typical for an interface with Pt. Hence, the presence of a Pt 6 nm coating is capable of severely breaking this otherwise perfect symmetry and hence of bringing about an additional undesirable non-reciprocity that must be regarded in the context of this article as an artefact. To avoid any ambiguity in interpretation of our experimental results, we have undertaken numerical simulations in order to estimate this unwanted contribution. These results are presented in Fig.2. More specifically, we have plotted the frequency difference $\Delta F$ = F($k_{sw}$) - F(-$k_{sw}$) accounting for non-reciprocity of SW propagation as a function of $k_{sw}$ for conventional NUSA pinning of one of the interfaces, which corresponds to the most undesirable situation.

As follows from this figure and as pointed out earlier, the almost negligible conventional non-reciprocity in thin films (2 nm and 4 nm) can be regarded as non-interfering experimental results even for the maximum $k_{sw}$ value of approximately $2{\cdot}10^5$ cm$^{-1}$ ($\Delta F$ = 10 MHz and 35 MHz respectively). In this sense the 2 nm film should be an ideal candidate. Sputtering technology used for film preparation proved to be inadequate for this extra-thin film; its unacceptably high inhomogeneity had been revealed though preliminary FMR characterization.

These preliminary estimations, both experimental and theoretical, have allowed us optimizing the conditions of the major experiment, consisting in revealing IDMI induced non-reciprocity and evaluating the constant D describing the strength of the IDMI pinning. Thus, reliably reproducible results have been obtained on the 4 nm sample. They are represented in Fig.3. More specifically, in Fig.3a)b)c) are given BLS spectra of the DE mode for $k_{sw}$ = 20.45 μm$^{-1}$, 11.81 μm$^{-1}$ and 4.1 μm$^{-1}$ corresponding to angles of incidence $\theta$ = 60°, 30° and 10°. The studied effect is small and to double-check the measured value of non-reciprocity we have taken measurements, for each value of $k_{sw}$, for two opposite values of the applied field $H$. In our case the magnetic field value has been chosen equal $H=$ ±1 kOe. We kept the field at this small level to make the DE mode dispersion as pronounced as possible, thus being more sensible to pinning. The evolution of the asymmetry can be easily traced; it is clearly visible in Fig.3a becomes less pronounced in Fig.3b and is practically inexistent in Fig.3c. This tendency is illustrated in more detail in Fig.4, where the dependence of the frequency shift $\Delta F$ (in GHz) due to IDMI for $k_{sw}$ > 0 (black solid line for theory and points for experiment) and $k_{sw}$ < 0 (red solid line for theory) is shown. The best agreement between theory and experiment has been attained with the IDMI constant D = 1.2 mJ/m$^2$, which is of realistic order of magnitude [3,23]. In dashed line are given theoretical estimations of a similar non-reciprocity due to conventional surface anisotropy in the configuration when it is most pronounced, namely the entirely asymmetric case with only one interface pinned. There can be no doubt that this undesirable in our case effect is too small to undermine the reliability of our observations.

This point which is vitally important for understanding of the obtained results, both experimental and theoretical, is schematically illustrated in Fig.5. More specifically, in the upper panels a) and b) are shown SW profiles in the case of conventional one-sided (only spins at the upper Py/Pt interface pinned) NUSA pinning in a) a thick and b) in a thin film. The conventional pinning is independent of the $k_{sw}$ sign and thus produces the same effect on the profiles of the two SWs propagating in opposite directions (green for $+k_{sw}$ and red for -$k_{sw}$), namely a characteristic drop of the dynamic magnetization in the vicinity of the pinned upper interface. In thicker films $2k_{sw}d_{Py}$ becomes comparable to 1 which results in a classical asymmetric exponential DE profiles. Importantly, the one corresponding to $+k_{sw}$ has its maximum at the upper interface while the other one has its minimum. It is not surprising that one-side pinning leads to two different profiles, hence non-negligible non-reciprocity. In thin films (Fig.5b) $2k_{sw}d_{Py}$ << 1 and, consequently the DDI dominated DE profile is practically uniform (Fig.5b). As a result, the red and green profiles are identical, which explains absence of any non-reciprocity.

By contrast, one-sided IDMI pinning changes its sign with inversion of the $k_{sw}$ which, when applied to an otherwise perfectly uniform DE profile in a thin film (Fig.5c), renders both profiles (red and green) asymmetric, but in different way. It is a characteristic drop in dynamic magnetization near the pinned surface for $+k_{sw}$ and no less characteristic rise for $-k_{sw}$ hence the non-reciprocity.

Reduction of the IDMI non-reciprocity with increase of the film thickness is natural. Our measurements have shown that the non-reciprocity $\Delta F$ for maximum value of $k_{sw}$ = 20.45 $\mu m^{-1}$, corresponding to θ = 60°, for the 6 nm film is approximately equal to 150 MHz, while that for the film 10 nm thick practically vanishes. It is not linear in $d_{Py}$, as predicted theoretically, which can be attributed to some variation in interface quality from sample to sample. Interestingly, no conventional non-reciprocity is observed for the thickest 10 nm sample, although it is large enough to be detected, as show our numerical estimations (see Fig.2). This can mean only one thing, namely that conventional pinning is distributed more or less symmetrically between two interfaces and thus it does not influence in any way the precision of the measurement of the effect investigated, i.e. IDMI induced non-reciprocity.

Another interesting feature related to surface pinning is revealed through the distortion of the dispersion curve (see Fig.6). The classical DE formula [24], derived in the magnetostatic approximation, is incapable of reproducing correctly the experimentally observed dispersion.

This is a clear indication that the wave of magnetization propagating in the investigated structure is of a more complex dipole-exchange nature [25], the exchange contribution originating from surface pinning. This has been fully confirmed by our numerical calculations, presented in Fig.6. Here we have concentrated on the overall run of the dispersion curve tracing an average value $(F(+k_{sw}) + F(-k_{sw}))/2$, both experimental and theoretical. It has turned out that it is the NUSA type pinning with the value found experimentally earlier that provides for excellent agreement. In other words, while conventional surface anisotropy appreciably effects general dispersion features, converting the negative curvature of the otherwise purely magnetostatic DE mode into a slightly positive. At the same time, as it has already been shown before the IDMI induced pinning leads to the splitting of the dispersion curve.

To conclude, we have experimentally revealed non-reciprocity of SW propagation in Py/Pt nanometric bi-layers induced by IDMI by means of Brillouin spectroscopy. Theoretical analysis has revealed that the observed non-reciprocity corresponds to the following value of the effective IDMI constant $D$ = 1.2 mJ/m$^2$. Although relatively low, it does not contradict earlier reported figures obtained by different techniques.

FIGURE CAPTIONS

Fig.1. Effective magnetization as a function of the inverse of the ferromagnetic film thickness $d_{Py}$.

Fig.2. Frequency difference between two spin waves propagating in opposite directions in Permalloy films of varying thickness, engendered by asymmetric one-side conventional pinning, as a function of their wave number. The normal uniaxial easy-axis surface anisotropy on the pinned interface is set 0.45 erg/cm$^2$.

Fig.3. BLS intensity as a function of the SW frequency for three characteristic angles of incidence: a) $k_{sw}$ =20.45 μm$^{-1}$, b) $k_{sw}$= 11.81 μm$^{-1}$ and c) $k_{sw}$ =4.1 μm$^{-1}$.

Fig.4. Frequency difference $\Delta F$ (in GHz) corresponding to IDMI induced non-reciprocity; theory versus experiment. The dashed lines respresent the theoretical corresponding values due the surface anisotropy with $k_{surf}$ =0.45 erg/cm$^2$.

Fig.5. Asymmetry of SW profiles and its role in shaping SW non-reciprocity.

Fig.6. Effect of surface anisotropy on SW dispersion. Theoretical and experimental mean values are given by a green solid line and filled black squares, respectively. Dashed lines (blue and red) indicate splitting due to IDMI.

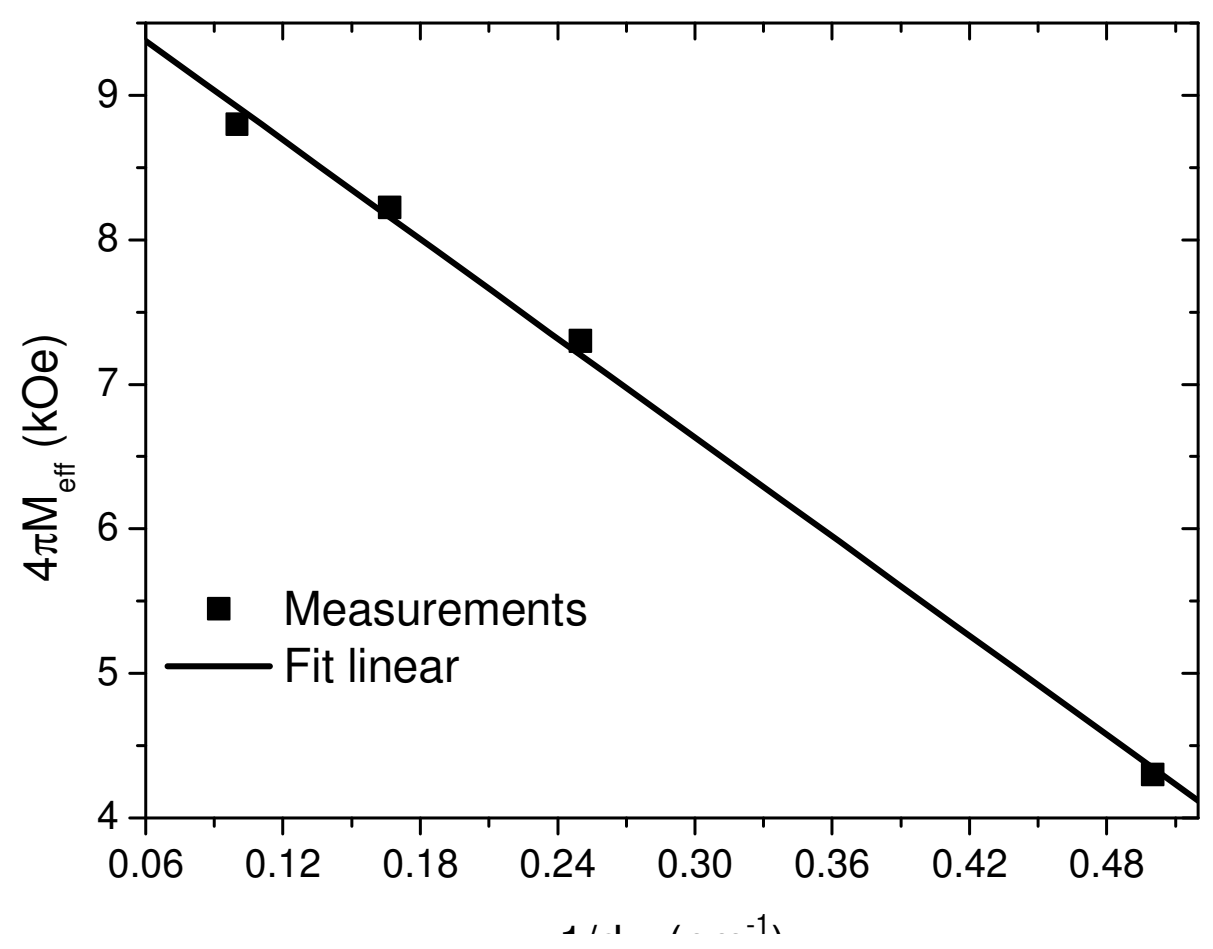

FIG.1.

A.A.Stashkevich, Phys. Rev. Lett.

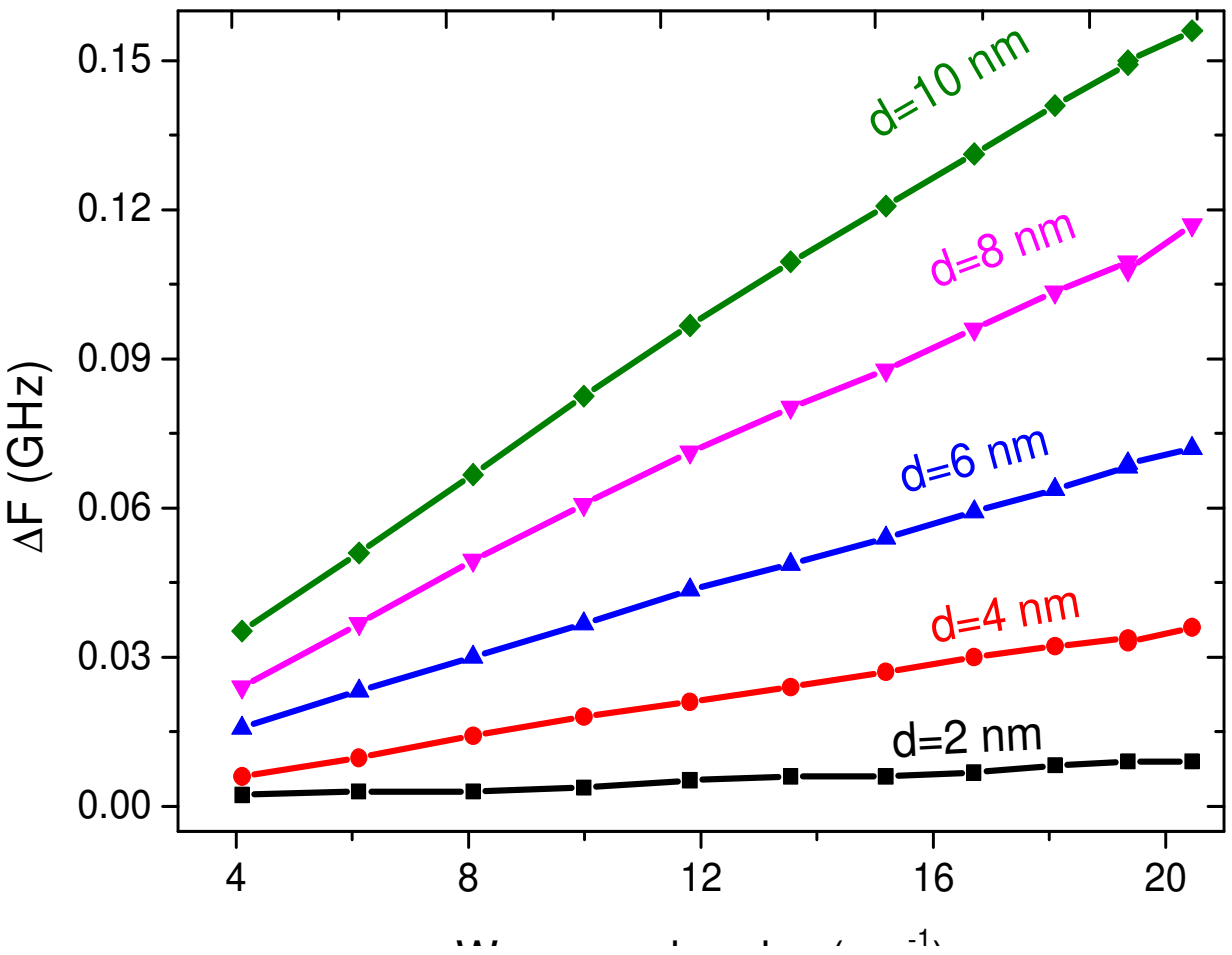

FIG.2.

A.A.Stashkevich, Phys. Rev. Lett.

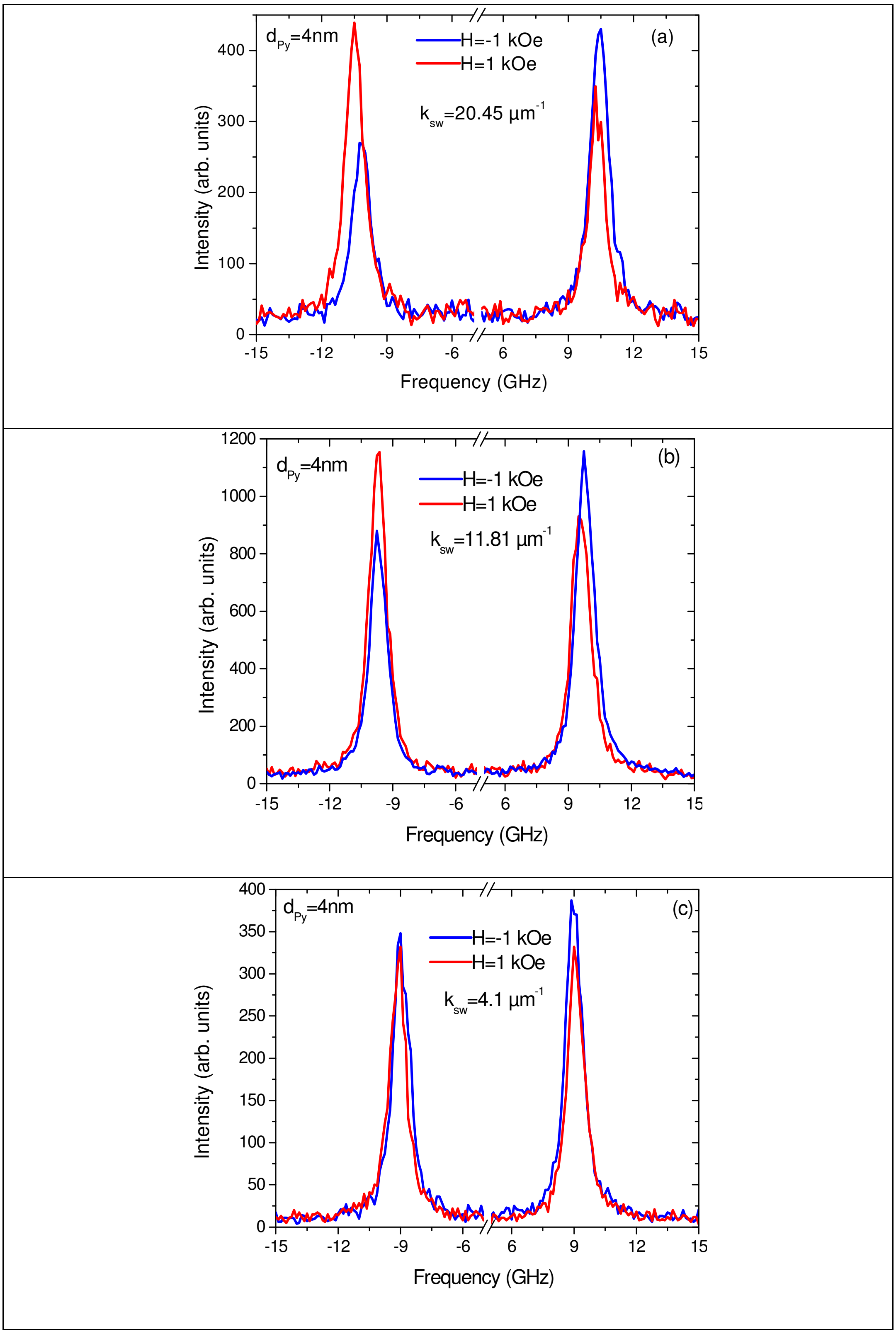

FIG.3.

A.A.Stashkevich, Phys. Rev. Lett.

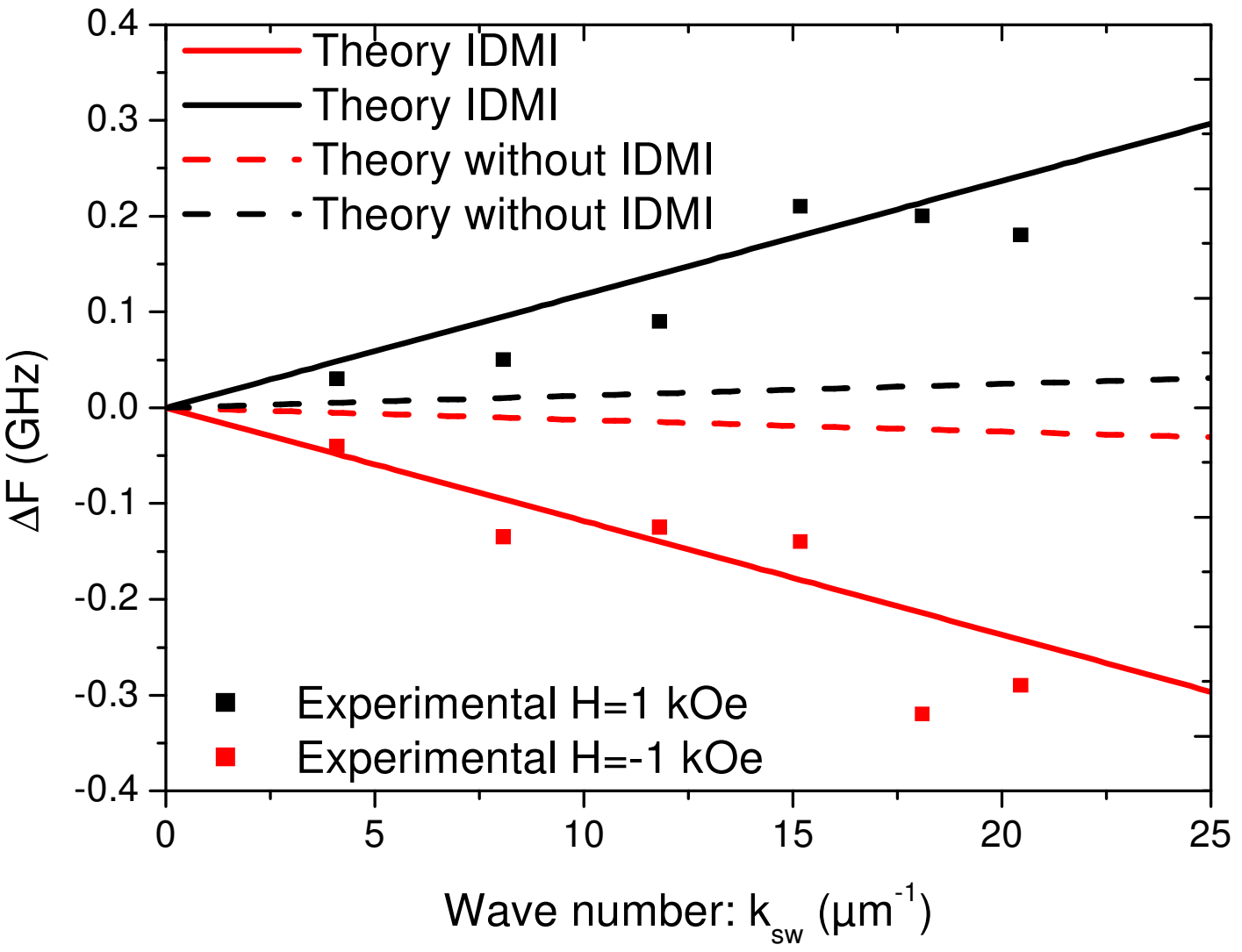

FIG.4.

A.A.Stashkevich, Phys. Rev. Lett.

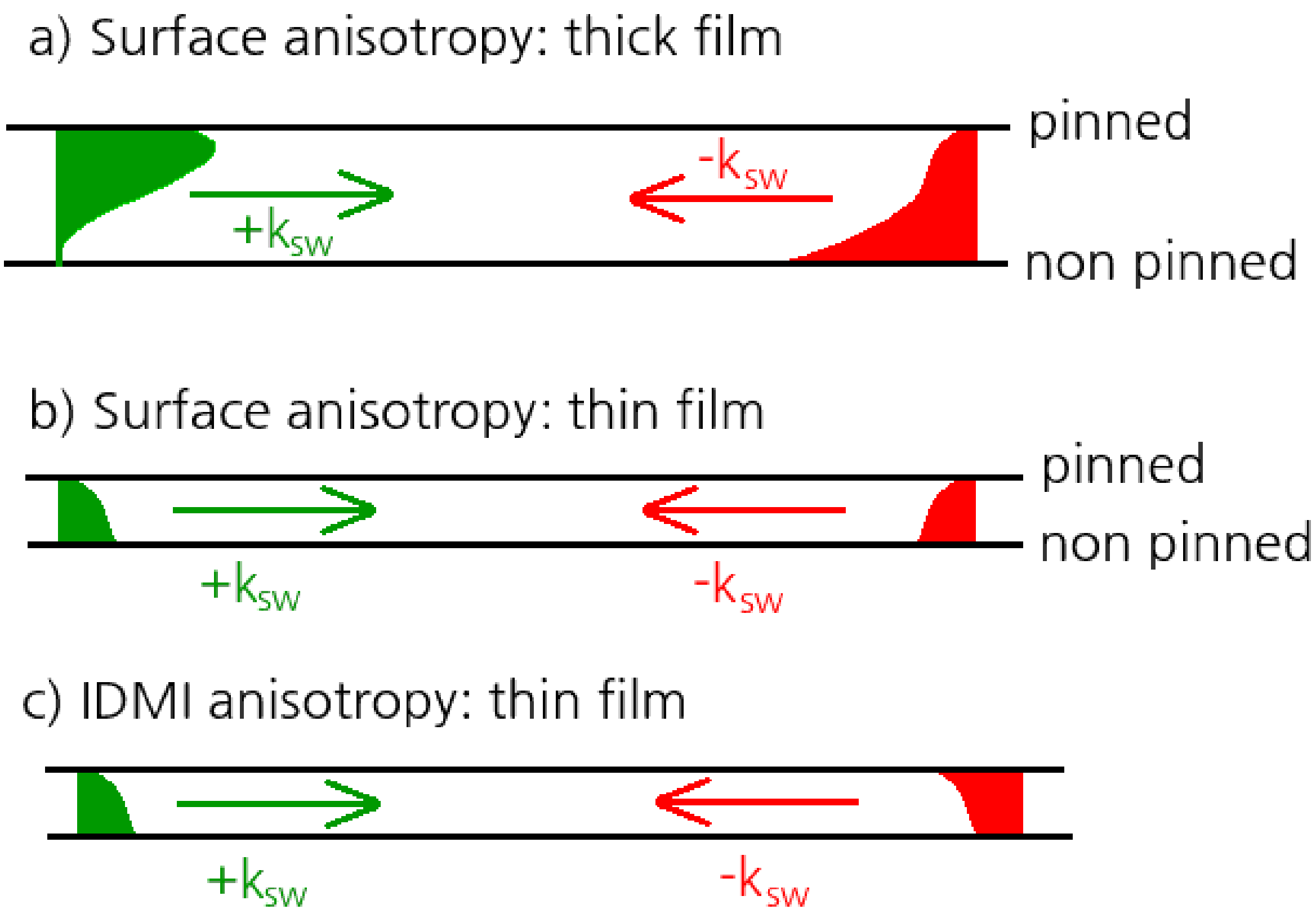

FIG.5.

A.A.Stashkevich, Phys. Rev. Lett.

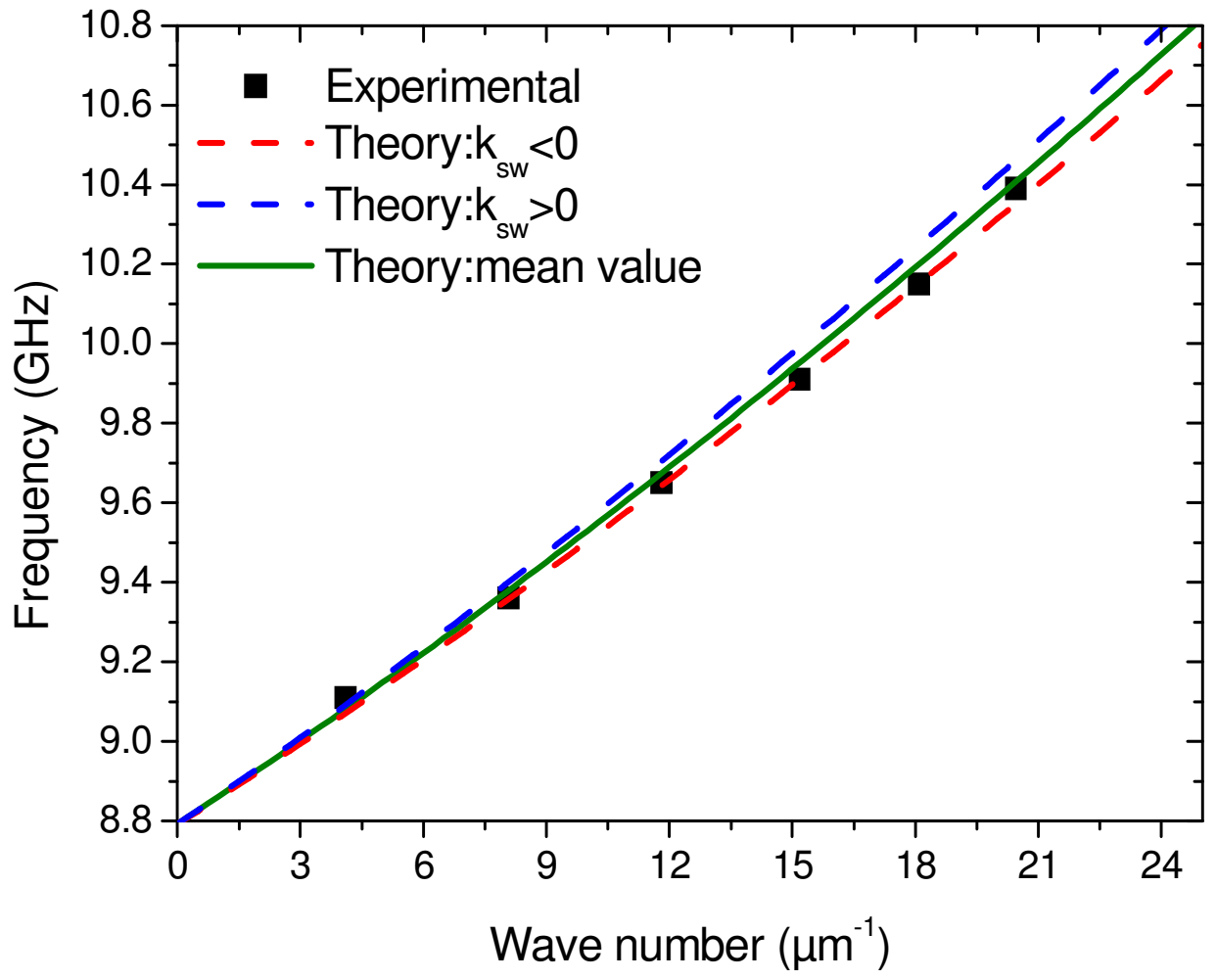

FIG.6.

A.A.Stashkevich, Phys. Rev. Lett.